\PassOptionsToPackage{unicode}{hyperref}
\PassOptionsToPackage{hyphens}{url}
\documentclass[
]{article}
\usepackage{amsmath,amssymb}
\usepackage{lmodern}
\usepackage{iftex}
\ifPDFTeX
  \usepackage[T1]{fontenc}
  \usepackage[utf8]{inputenc}
  \usepackage{textcomp} 
\else 
  \usepackage{unicode-math}
  \defaultfontfeatures{Scale=MatchLowercase}
  \defaultfontfeatures[\rmfamily]{Ligatures=TeX,Scale=1}
\fi
\IfFileExists{upquote.sty}{\usepackage{upquote}}{}
\IfFileExists{microtype.sty}{
  \usepackage[]{microtype}
  \UseMicrotypeSet[protrusion]{basicmath} 
}{}
\makeatletter
\@ifundefined{KOMAClassName}{
  \IfFileExists{parskip.sty}{%
    \usepackage{parskip}
  }{
    \setlength{\parindent}{0pt}
    \setlength{\parskip}{6pt plus 2pt minus 1pt}}
}{
  \KOMAoptions{parskip=half}}
\makeatother
\usepackage{xcolor}
\IfFileExists{xurl.sty}{\usepackage{xurl}}{} 
\IfFileExists{bookmark.sty}{\usepackage{bookmark}}{\usepackage{hyperref}}
\hypersetup{
  pdftitle={Agile (data) science: a (draft) manifesto},
  pdfauthor={J. J. Merelo},
  hidelinks,
  pdfcreator={LaTeX via pandoc}}
\usepackage[margin=1in]{geometry}
\usepackage{graphicx}
\makeatletter
\def\maxwidth{\ifdim\Gin@nat@width>\linewidth\linewidth\else\Gin@nat@width\fi}
\def\maxheight{\ifdim\Gin@nat@height>\textheight\textheight\else\Gin@nat@height\fi}
\makeatother
\setkeys{Gin}{width=\maxwidth,height=\maxheight,keepaspectratio}
\makeatletter
\def\fps@figure{htbp}
\makeatother
\newlength{\cslhangindent}
\newlength{\csllabelwidth}
\newlength{\cslentryspacingunit} 
\newenvironment{CSLReferences}[2] 
 {
  \setlength{\parindent}{0pt}
  \ifodd #1
  \let\oldpar\par
  \def\par{\hangindent=\cslhangindent\oldpar}
  \fi
  \setlength{\parskip}{#2\cslentryspacingunit}
 }%
 {}
\usepackage{calc}

\ifLuaTeX
  \usepackage{selnolig}  
\fi

\title{Agile (data) science: a (draft) manifesto}
\author{J. J. Merelo}
\date{4/7/2022}

\begin{document}
\maketitle
\begin{abstract}
Science has a data management problem, as well as a project management
problem. While industrial-grade data science teams have embraced the
\emph{agile} mindset, and adopted or created all kind of tools to create
reproducible workflows, academia-based science is still (mostly) mired
in a mindset that is focused on a single final product (a paper),
without focusing on incremental improvement, on any specific problem or
customer, or, paying any attention reproducibility. In this report we
argue towards the adoption of the agile mindset and agile data science
tools in academia, to make a more responsible, and over all,
reproducible science.
\end{abstract}

\hypertarget{introduction}{%
\subsection{Introduction}\label{introduction}}

By \emph{agile}, we usually imply a mindset that is applied to the whole
software development lifecycle which is customer-centered and focused on
continuous improvement of increasingly complex minimally viable
products. The name comes from the Agile Manifesto (Beck et al. 2001),
literally ``Manifesto for agile software development''. This manifesto
has certainly changed the way software at large is developed, and become
mainstream, spawning many different methodologies and best practice
guidelines. It has proved to be an efficient way of carrying out all
kind of projects, from small to large-scale ones, mitigating the
presence of bugs and proving to be more efficient (Abrahamsson et al.
2017) than the methodology that prevailed previously (and still today in
many sectors), generally called \emph{waterfall} (Andrei et al. 2019),
which separated (or \emph{siloed}) different teams doing from the
specification to the testing, with every team acting at different parts
of the lifecycle.

Despite being prevalent in software development (and, in general,
project development) environments, it certainly has not reached science
at large, which arguably follows a method that closely follows the
waterfall methodology.

Since data science and engineering has become an integral part of the
workflow in many companies, \emph{agile} data science is, mostly, the
way it's done. Again, this is mostly because data science is mostly done
in the industry, and not in academia, which does not have the same kind
of workflows to deal with its own data.

Our intention is to try and put science back in data science. We will
try and examine critically how science is done, what are the main
reasons why this agile mindset is not being used in science, how would
\emph{agile} concepts translate to science, and eventually what agile
data science and science at large woudl look like.

We will first present what attempts have been made to translate agile
concepts to the (academic) world of science.

\hypertarget{state-of-the-art}{%
\subsection{State of the art}\label{state-of-the-art}}

Despite being an age-old pursuit and probably, as such, ripe for
disruption, science has been done in pretty much the same way for years.
In very rough brushstrokes, it starts with application for funding, that
includes a workplan, that generally works hierarchically from principal
investigators to senior and then junior researchers, producing a series
of artifacts which always include papers (which are snapshots of the
state of the art), and in some cases software, protocols, or even, in
some limited areas (mainly astronomy and medicine), workflows.

This situation has been challenged repeatedly, lately, mainly after the
introduction of the aforementioned agile manifesto. In (Amatriain and
Hornos 2009), which is essentially a presentation and not a formal
paper, several proposal are made to apply agile ``methods'' in research;
something that has been proposed repeatedly in later years, for instance
in this blog post (Carattino, n.d.) and even in this paper (Baijens,
Helms, and Iren 2020) which specifies an agile methodology, Scrum, and
how it can be applied specifically to data science projects. As a matter
of fact, there were several attempts to raise the issue again and bring
it to the attention of the research community: a blog post introduced
\emph{agile research} (Amatriain 2008) and even drafted an Agile
Research Manifesto (Amatriain 2009). This was almost totally forgotten
until it was brought up two years ago in a blog called ``Agile Science''
(Bergman 2018). Independently, some researchers proposed an (almost)
ultimatum for Agile Research in (Way, Chandrasekhar, and Murthy 2009),
and eventually it became fruitful in a restricted environment, mHealth,
in (Wilson et al. 2018). This goes to show that it's still part of the
fringe, and has not been incorporated either to funding agencies
guidelines, or to the common science and research practice.

This is certainly related with Open Science: Open Science adapts the
main ideas of open source software development to the publication of
scientific results and artifacts; the push for Open Science (Robson et
al. 2021) has provided with new venues and new ways of understanding and
producing science. However, the uptake of new methodologies is still
very slow. While most companies have created pipelines for data
management (Rodríguez 2019), there are neither clear guidelines or best
practices nor resources where scientific data management can be done at
scale and, what's more important, in a way that can have a (positive)
outcome for your career.

\hypertarget{hypothesis-on-agile-science}{%
\subsection{Hypothesis on agile
science}\label{hypothesis-on-agile-science}}

We certainly need to acknowledge first that science, the way it is now,
has many problems. Many of them start and end with funding, but we
should also realize that using XIX century formats to publish XXI
century research leaves a lot to be desired. However, the fundamental
problem is not in products, is in practitioner's workflows themselves,
and this ends in frustration and, when major crisis like COVID-19
strike, major problems carrying out much needed research that can be
used as a foundation for the next, also necessary, step. So we will try
to present, and defend, a series of hypothesis that would be the
foundation of agile science, and that would contribute to solve the data
management problem science currently has.

\hypertarget{reproducibility-over-publishability}{%
\subsubsection{Reproducibility over
publishability}\label{reproducibility-over-publishability}}

Science arguably can't be science if it's not reproducible. However,
there are many practical hurdles for it to be so. First, the ``paper''
format, even if it's paper-with-embedded-links, is not reproducible
\emph{per se} (even if tools such as the one proposed by Kardas et al.
(Kardas et al. 2020) are going to be able to extract at least the
\emph{results} in papers, not the code and experimental setup). Efforts
like \emph{Papers with Code} ({``Papers with Code,''} n.d.) help
associate something that has been already published with its
corresponding code, but even if this is a step forward reproducibility,
a effort similar to the one invested in deploying applications to
production should be made: configuration as code, provision of all
needed services, inclusion of all data inflows, as well as extensive
testing. Testing that is an essential part of the agile mindset, and is
severely lacking in scientific workflows.

\hypertarget{testing-at-all-levels-over-hypotheses-proved-once}{%
\subsubsection{Testing at all levels over hypotheses proved
once}\label{testing-at-all-levels-over-hypotheses-proved-once}}

Essentially, all papers try to prove something, to answer a research
question or to prove a hypothesis over data that is characterized in a
certain way. Datasets are static, and hypotheses are proved over
provided datasets, which are increasingly available, although that is
not necessarily a given.

Science practitioners know that working with raw data is always hard,
and needs a series of steps to be suitable for use. Extracting the data
is hard, checking that everything is correct is hard. Small changes
could lead to invalidation of results.

This is why testing is essential. ``Code that's not tested is broken'',
we could say dataflows and workflows that are not tested are broken.
Besides basic testing (testing for duplication, invalid data, things
like that), we should formulate tests on data to check that it keeps
being in the same format, range and general characteristics that allow
our hypothesis to be valid. So agile science would need to test data to
start with, before using it as input for workflows, but it should also
unit-test all software used, perform integration tests on data +
software, and eventually transform the hypothesis into actual software
tests that would \emph{continuously} check if the hypothesis still
holds.

Increasingly, and when the Internet itself, as well as myriad sensors
and devices, is a continuous source of data, publishing a paper drawing
conclusions over a small piece of data is valuable and helpful. Creating
a tested, continuously deployed workflow that over and over again tests
that hypothesis, and that has been released as free software to be
integrated as input or middleware in other workflows is immensely more
valuable. But needs another hypothesis

\hypertarget{open-over-closed}{%
\subsubsection{Open over closed}\label{open-over-closed}}

Science should be reproducible, and this implies that software used or
produced for it should be open too. However, it makes sense to emphasize
the pliability and flexibility of free-software-as-science. If you want
to configure increasingly complicated workflows that are going to be
deployed continuously, a single non-open component would break them and
make them impossible.

I would like to think this is the least controversial hypothesis that
would support agile science. After all, it's been repeatedly proved
{[}Vandewalle (2012){]}(Vandewalle 2019) that papers with code have a
higher impact than those who hide it or simply don't publish it
alongside the paper itself. As a matter of fact, it's quite usual right
now in many scientific conferences to accompany the paper with pointers
to a GitHub repo; searching over GitHub for similar code might also help
scientists/coders as much as reading new papers. Some important
conferences like NeurIPS now have reproducibility responsibles in their
program committee, and they have shown that 3 out of 4 papers in that
conference already post their GitHub repository (Gibney 2020). This is
mainly prevalent, however, in the data science/machine learning
community (Wattanakriengkrai et al. 2020).

There's a more important aspect to this: science in open repositories
and with free licenses leaves research open to all stakeholders, who can
have a say in its outcomes, as well as in their direction. Which is why
we prefer:

\hypertarget{stakeholder-collaboration-over-vertical-chains-of-command}{%
\subsubsection{Stakeholder collaboration over vertical
chains-of-command}\label{stakeholder-collaboration-over-vertical-chains-of-command}}

Again, it has been the Covid crisis the occasion where the world has
realized that it needed science, it needed a lot and it needed it now.
The whole world health and even lives were at stake, and government
officials as well as the civil society needed to know from what to do to
avoid contagion to the evolution of the pandemic and what it would mean
vis-à-vis their return to a less restrictive situation.

In this, many open data repositories emerged with partial, localized,
solutions, that allowed people in general and maybe also government
officials have a bit of more control over their lives.

As a matter of fact, the scientific chain on command and general career
environment causes not a few problems in participants (Levecque et al.
2017) . This study mentions explicitly ``the supervisor's leadership
style'' as one of the factors that are linked to health problems, with
the general work and organizational context cited as one of the
highlights of the study. The scenario shown in a recent Nature survey
(Woolston 2019) would be unsustainable in any kind of company, be it
software development or any other kind of company. Agile software
development offers a series of principles and best practices that enable
a sustainable pace of development, with regular deployments; but since
it values ``individuals and interactions over processes and tools'' it
also creates an humane, self-organizing work environment that results in
better worker satisfaction, as well as sustainable, career-long
learning. That should also be an objective in scientific research.

Techniques such as Scrum (Baijens, Helms, and Iren 2020) have helped in
data science (and this, only recently); again, this has not extended to
computer science or, in general, science at large.

\hypertarget{the-way-forward}{%
\subsection{The way forward}\label{the-way-forward}}

Agile fixed software development by proposing a series of principles
attached to the Agile manifesto (Beck et al. 2001) that eventually
spawned a series of tools on one hand, and best practices in other hand.
Tools that can be grouped into generic CI and CD toolchains, including
MLOps tools Kreuzberger, Kühl, and Hirschl (2022) , team work tools
(usually attached to source repositories) such as Jira or GitHub itself
and the use of different methodologies (Kanban (Ahmad, Markkula, and
Oivo 2013), Scrum), practices (reviews, retrospective meetings) and
roles (product owner, stakeholder) to streamline software production,
bring value to stakeholders, and provide a sane, stable nurturing and
eventually productive working environment. I have been advocating for
using these techniques for quite a long time, at least since 2011 (last
version of the talk on ``the art of evolutionary algorithm programming
is here (Merelo 2013)). In this report I try to put everything together
under the same framework which we will be calling agile science.

Science should not be different, and a (roughly) direct translation of
all these practices, however they are interpreted, would be beneficial.
We'll try, anyway, to delve a bit further into those concepts to see how
they translate and how they could be applied, in practice, to science.

\hypertarget{the-product-needs-to-be-a-deployed-workflow}{%
\subsubsection{The product needs to be a deployed
workflow}\label{the-product-needs-to-be-a-deployed-workflow}}

We need to shift focus from publishing on paper as a one shot to
deploying working workflows, which will have as side products papers
that can be continuously updated or else simply remain as a snapshot of
the state of the art in a particular point in time, but that can,
anyway, be re-produced (see one of the hypothesis above, reproducibility
above all else) by anyone, but specially the people that have produced
it first.

As we see this shift in action, we will see how the science system
changes to accommodate this and take it into account in a scientific
career. At this point in time, the peer-review and article-publishing
system has been gamed in so many ways (Arney, n.d.), that it's almost
meaningless to rely on it for scientific promotion, not to mention
science itself. To a certain point, it's become an obstacle to science.
But replicability can fix it (Moonesinghe, Khoury, and Janssens 2007),
and the best way to totally replicate results is to publish them openly
(third hypothesis above).

There are clearly lots of hurdles to overcome in this, including the
fact that scientific publishing powerhouses will have to become workflow
hosting players. That, however, is an externality to this proposal and
can no doubt be solved as soon as the economic scenario draws itself.

\hypertarget{the-product-owner-would-be-person-with-the-original-idea}{%
\subsubsection{The product owner would be person with the original
idea}\label{the-product-owner-would-be-person-with-the-original-idea}}

The 4th hypothesis above advocates for stakeholder participation over
hierarchical decision-making. However, in agile teams there must be a
person that will be \emph{calling the shots}. A successful product owner
should be able to (Oomen et al. 2017) to ``define the product vision''.
Since in this case the \emph{product} is a scientific workflow, the
person that had the original idea should be the one that \emph{owns} the
product, and should them prioritize different paths of development,
define minimally viable products and its consequent milestones, and in
general, take not only responsibility for the finished product but also
work with the team to achieve success.

That sense of ownership will one of the factors decreasing anxiety and
increasing sustainable productivity in science. It dodges hierarchical
organization by making the principal investigator product owner only if
he or she, effectively, has had the original idea and the vision of
taking it forward. At the same time, a product owner is part of an agile
team, and \emph{owns} the product, not the team. This small semantic
change also simplifies roles and clarifies what everyone will be doing
in the product development lifecycle.

\hypertarget{use-common-software-development-tools-and-practices}{%
\subsubsection{Use common software development tools and
practices}\label{use-common-software-development-tools-and-practices}}

Most scientific development includes, or even is just simply, software
development. We should use common software development tools and best
practices, and integrate them into the development of the scientific
workflow that is ultimately the objective.

In many cases, and specially in data science/machine learning, there
will specialized tools such as MLflow (Zaharia et al. 2018) with
frontends such as Snapper ML (Domenech and Guillén 2020) to simplify the
creation of workflows. No doubt these workflow high-level tools will be
extended to other fields, and integrated with mainstream deployment
tools such as Docker or Kubernetes. Integrating these practices
seamlessly merges product (workflow) development with software
development, and also leverages existing tools such as GitHub or GitLab
with their accompanying workflow design tools (Github Actions,
pipelines), as well as other cloud environments with their accompanying
tools.

This also decouples the production of a workflow from its actual
deployment. As long as deployment is clearly expressed, it can be
deployed by the scientific team producing it on premises or on the
cloud, or done by anyone else elsewhere. One could even think about
global free infrastructure for doing this kind of thing, or even a model
similar to pay-to-publish: pay-to-deploy, and let the hosting place take
care of long-term maintenance. This also makes science and the
scientific effort, much more sustainable, and satisfies stakeholders
(fourth hypothesis above) by keeping the product of science funding
available and working way beyond the mere existence of the grant, or
even the group itself.

\hypertarget{conclusions}{%
\subsection{Conclusions}\label{conclusions}}

In this report we have tried to propose a set of best practices that we
think would benefit science at large, but especially those disciplines
that rely heavily in data and software to produce results. Essentially,
it interprets, translates and codifies the agile (software development)
manifesto to the scientific arena, converting what this manifesto values
in a series of 4 hypotheses that will guide agile science.

Hypotheses need to be proved, however, and science prides itself in
being able to establish \emph{fact} over anecdotal, or even
counter-intuitive, evidence. This is why we also provide a \emph{path
forward} in the shape of several best practices suggestions that will
help prove those hypotheses beyond any doubt. There is strong evidence
that supports them, and our own experience using it for some time via
open-repository product development, specially in papers such as
(García-Ortega, Sánchez, and Merelo-Guervós 2021) and, in general, most
papers that we have published lately, helps stakeholder participation in
the production of workflows, makes easier to evolve software related to
science and streamline product development, and makes also easier to
respond to new requirements. Proving the positive effects of these
\emph{preferences} is, however, left as future work.

\hypertarget{acknowledgements}{%
\subsection{Acknowledgements}\label{acknowledgements}}

This research was funded by projects TecNM-5654.19-P and DemocratAI
PID2020-115570GB-C22.

\hypertarget{references}{%
\subsection*{References}\label{references}}
\addcontentsline{toc}{subsection}{References}

\hypertarget{refs}{}
\begin{CSLReferences}{1}{0}
\leavevmode\vadjust pre{\hypertarget{ref-abrahamsson2017agile}{}}%
Abrahamsson, Pekka, Outi Salo, Jussi Ronkainen, and Juhani Warsta. 2017.
{``Agile Software Development Methods: Review and Analysis.''}
\url{https://arxiv.org/abs/1709.08439}.

\leavevmode\vadjust pre{\hypertarget{ref-ahmad2013kanban}{}}%
Ahmad, Muhammad Ovais, Jouni Markkula, and Markku Oivo. 2013. {``Kanban
in Software Development: A Systematic Literature Review.''} In
\emph{2013 39th Euromicro Conference on Software Engineering and
Advanced Applications}, 9--16. IEEE.

\leavevmode\vadjust pre{\hypertarget{ref-AgileResearchBlog}{}}%
Amatriain, Xavier. 2008. {``Agile Research.''}
\url{http://technocalifornia.blogspot.com/2008/06/agile-research.html}.

\leavevmode\vadjust pre{\hypertarget{ref-AgileResearch}{}}%
---------. 2009. {``A Manifesto for Agile Research.''}
\url{https://xamat.github.io/AgileResearch/}.

\leavevmode\vadjust pre{\hypertarget{ref-agile-science}{}}%
Amatriain, Xavier, and Gemma Hornos. 2009. {``Agile Methods in
Research.''} \url{https://www.slideshare.net/xamat/agile-science}.

\leavevmode\vadjust pre{\hypertarget{ref-andrei2019study}{}}%
Andrei, Bogdan-Alexandru, Andrei-Cosmin Casu-Pop, Sorin-Catalin
Gheorghe, and Costin-Anton Boiangiu. 2019. {``A Study on Using Waterfall
and Agile Methods in Software Project Management.''} \emph{Journal Of
Information Systems \& Operations Management}, 125--35.

\leavevmode\vadjust pre{\hypertarget{ref-arneybroken}{}}%
Arney, Kat. n.d. {``Science Is Broken. Here's How to Fix It.''}
\url{http://littleatoms.com/science/science-broken-heres-how-fix-it}.

\leavevmode\vadjust pre{\hypertarget{ref-9140255}{}}%
Baijens, J., R. Helms, and D. Iren. 2020. {``Applying Scrum in Data
Science Projects.''} In \emph{2020 IEEE 22nd Conference on Business
Informatics (CBI)}, 1:30--38.
\url{https://doi.org/10.1109/CBI49978.2020.00011}.

\leavevmode\vadjust pre{\hypertarget{ref-beck2001manifesto}{}}%
Beck, Kent, Mike Beedle, Arie Van Bennekum, Alistair Cockburn, Ward
Cunningham, Martin Fowler, James Grenning, et al. 2001. {``Manifesto for
Agile Software Development.''}

\leavevmode\vadjust pre{\hypertarget{ref-XavierAgileScience}{}}%
Bergman, Olle. 2018.
\url{https://crastina.se/xavier-invented-agile-science-a-decade-ago/}.

\leavevmode\vadjust pre{\hypertarget{ref-carattino-agile-dev}{}}%
Carattino, Aquiles. n.d. {``Agile Development for Science: Scientific
Work Can Also Benefit from Principles Derived from Software
Development.''}
\url{https://www.uetke.com/blog/general/agile-development-for-science/}.

\leavevmode\vadjust pre{\hypertarget{ref-Molner_Domenech_2020}{}}%
Domenech, Antonio Molner, and Alberto Guillén. 2020. {``Ml-Experiment: A
Python Framework for Reproducible Data Science.''} \emph{Journal of
Physics: Conference Series} 1603 (September): 012025.
\url{https://doi.org/10.1088/1742-6596/1603/1/012025}.

\leavevmode\vadjust pre{\hypertarget{ref-garciaortega2021tropes}{}}%
García-Ortega, Rubén Héctor, Pablo García Sánchez, and Juan J.
Merelo-Guervós. 2021. {``Tropes in Films: An Initial Analysis.''}
\url{https://arxiv.org/abs/2006.05380}.

\leavevmode\vadjust pre{\hypertarget{ref-gibney2020ai}{}}%
Gibney, Elizabeth. 2020. {``This AI Researcher Is Trying to Ward Off a
Reproducibility Crisis.''} \emph{Nature} 577 (7788): 14.

\leavevmode\vadjust pre{\hypertarget{ref-kardas2020axcell}{}}%
Kardas, Marcin, Piotr Czapla, Pontus Stenetorp, Sebastian Ruder,
Sebastian Riedel, Ross Taylor, and Robert Stojnic. 2020. {``Axcell:
Automatic Extraction of Results from Machine Learning Papers.''}
\emph{arXiv Preprint arXiv:2004.14356}.

\leavevmode\vadjust pre{\hypertarget{ref-kreuzberger2022machine}{}}%
Kreuzberger, Dominik, Niklas Kühl, and Sebastian Hirschl. 2022.
{``Machine Learning Operations ({MLOps}): Overview, Definition, and
Architecture.''} \emph{arXiv Preprint arXiv:2205.02302}.

\leavevmode\vadjust pre{\hypertarget{ref-Levecque2017WorkOA}{}}%
Levecque, K., F. Anseel, A. D. Beuckelaer, J. Heyden, and L. Gisle.
2017. {``Work Organization and Mental Health Problems in PhD
Students.''} \emph{Research Policy} 46: 868--79.

\leavevmode\vadjust pre{\hypertarget{ref-makinen2021needs}{}}%
Mäkinen, Sasu, Henrik Skogström, Eero Laaksonen, and Tommi Mikkonen.
2021. {``Who Needs MLOps: What Data Scientists Seek to Accomplish and
How Can MLOps Help?''} \emph{arXiv Preprint arXiv:2103.08942}.

\leavevmode\vadjust pre{\hypertarget{ref-merelo13}{}}%
Merelo, JJ. 2013. {``The Art of Evolutionary Algorithm Programming.''}
\url{https://issuu.com/jjmerelo/docs/art-ecp-cec13}.

\leavevmode\vadjust pre{\hypertarget{ref-moonesinghe2007most}{}}%
Moonesinghe, Ramal, Muin J Khoury, and A Cecile JW Janssens. 2007.
{``Most Published Research Findings Are False---but a Little Replication
Goes a Long Way.''} \emph{PLoS Med} 4 (2): e28.

\leavevmode\vadjust pre{\hypertarget{ref-oomen2017can}{}}%
Oomen, Sandra, Benny De Waal, Ademar Albertin, and Pascal Ravesteyn.
2017. {``How Can Scrum Be Succesful? Competences of the Scrum Product
Owner.''}

\leavevmode\vadjust pre{\hypertarget{ref-pwc}{}}%
{``Papers with Code.''} n.d. \url{https://paperswithcode.com}.

\leavevmode\vadjust pre{\hypertarget{ref-robson_opensci}{}}%
Robson, Samuel G, Myriam A Baum, Jennifer L Beaudry, Julia Beitner,
Hilmar Brohmer, Jason Chin, Katarzyna Jasko, et al. 2021. {``Nudging
Open Science.''} PsyArXiv. \url{https://doi.org/10.31234/osf.io/zn7vt}.

\leavevmode\vadjust pre{\hypertarget{ref-uberlyftothers}{}}%
Rodríguez, Jesús. 2019. {``{How LinkedIn, Uber, Lyft, Airbnb and Netflix
are Solving Data Management and Discovery for Machine Learning
Solutions}.''}
\url{https://www.kdnuggets.com/2019/08/linkedin-uber-lyft-airbnb-netflix-solving-data-management-discovery-machine-learning-solutions.html\#disqus_thread}.

\leavevmode\vadjust pre{\hypertarget{ref-vandewalle2012code}{}}%
Vandewalle, Patrick. 2012. {``Code Sharing Is Associated with Research
Impact in Image Processing.''} \emph{Computing in Science \&
Engineering} 14 (4): 42--47.

\leavevmode\vadjust pre{\hypertarget{ref-vandewalle2019code}{}}%
---------. 2019. {``Code Availability for Image Processing Papers: A
Status Update.''} In \emph{WIC IEEE SP Symposium on Information Theory
and Signal Processing in the Benelux, Date: 2019/05/28-2019/05/29,
Location: Gent, Belgium}.

\leavevmode\vadjust pre{\hypertarget{ref-wattanakriengkrai2020github}{}}%
Wattanakriengkrai, Supatsara, Bodin Chinthanet, Hideaki Hata, Raula
Gaikovina Kula, Christoph Treude, Jin Guo, and Kenichi Matsumoto. 2020.
{``GitHub Repositories with Links to Academic Papers: Open Access,
Traceability, and Evolution.''} \url{https://arxiv.org/abs/2004.00199}.

\leavevmode\vadjust pre{\hypertarget{ref-way2009agile}{}}%
Way, Thomas, Sandhya Chandrasekhar, and Arun Murthy. 2009. {``The Agile
Research Penultimatum.''} In \emph{Software Engineering Research and
Practice}, 530--36. Citeseer.

\leavevmode\vadjust pre{\hypertarget{ref-wilson2018agile}{}}%
Wilson, Kumanan, Cameron Bell, Lindsay Wilson, and Holly Witteman. 2018.
{``Agile Research to Complement Agile Development: A Proposal for an
mHealth Research Lifecycle.''} \emph{NPJ Digital Medicine} 1 (1): 1--6.

\leavevmode\vadjust pre{\hypertarget{ref-woolston2019phds}{}}%
Woolston, Chris. 2019. {``PhDs: The Tortuous Truth.''} \emph{Nature} 575
(7782): 403--7.

\leavevmode\vadjust pre{\hypertarget{ref-zaharia2018accelerating}{}}%
Zaharia, Matei, Andrew Chen, Aaron Davidson, Ali Ghodsi, Sue Ann Hong,
Andy Konwinski, Siddharth Murching, et al. 2018. {``Accelerating the
Machine Learning Lifecycle with MLflow.''} \emph{IEEE Data Eng. Bull.}
41 (4): 39--45.

\end{CSLReferences}

\end{document}